\documentclass[journal=jacsat,manuscript=article]{achemso}
\usepackage{chemformula} 
\usepackage[T1]{fontenc} 
\usepackage{graphicx,color}
\usepackage{amsfonts,amsmath,amssymb,amsbsy}
\usepackage[colorlinks=true, linkcolor=blue, citecolor=blue, urlcolor=blue]{hyperref}
\usepackage{float}

\author{Jiwen Chen}
\affiliation{College of Engineering Physics, Shenzhen Technology University, Shenzhen, 518118, China}

\author{Laichuan Shen}
\affiliation{The Center for Advanced Quantum Studies and Department of Physics, Beijing Normal University, 100875 Beijing, China}

\author{Hongyu An}
\affiliation{College of New Materials and New Energies, Shenzhen Technology University, Shenzhen, 518118, China}

\author{Xichao Zhang}
\affiliation{Department of Applied Physics, Waseda University, Okubo, Shinjuku-ku, Tokyo 169-8555, Japan}

\author{Hua Zhang}
\affiliation{College of Engineering Physics, Shenzhen Technology University, Shenzhen, 518118, China}

\author{Haifeng Du}
\affiliation{Anhui Province Key Laboratory of Low-Energy Quantum Materials and Devices, High Magnetic Field Laboratory, HFIPS, Chinese Academy of Sciences, Hefei, 230031, China}

\author{Xiaoguang Li}
\email{lixiaoguang@sztu.edu.cn}
\affiliation{College of Engineering Physics, Shenzhen Technology University, Shenzhen, 518118, China}

\author{Yan Zhou}
\email{zhouyan@cuhk.edu.cn}
\affiliation{School of Science and Engineering, The Chinese University of Hong Kong, Shenzhen, 518172, China}

\title{Magnetic Bimeron Traveling on the Domain Wall}

\begin{document}







\newpage
\begin{abstract}
  Domain wall bimerons (DWBMs) are nanoscale spin textures residing within the magnetic domain walls of in-plane magnets. In this study, we employ both numerical and analytical methods to explore the stabilization of Néel-type domain wall bimerons and their dynamics when excited by spin-orbit torque. Our findings reveal two unique and intriguing dynamic mechanisms, which depend on the polarization direction of the spin current: In the first scenario, the magnetic domain wall serves as a track that confines the motion of the bimeron and effectively suppresses the skyrmion Hall effect. In the second scenario, pushing the magnetic domain wall triggers a rapid sliding of the bimeron along the wall. This process significantly enhances the dynamics of the bimeron, resulting in a velocity increase of approximately 40 times compared to skyrmions and bimeron solitons. Our results highlight the potential advantages of the skyrmion Hall effect in developing energy-efficient spintronic devices based on domain wall bimerons.

\end{abstract}
\vspace{5mm}

 A bimeron can be regarded as a counterpart of the skyrmion in in-plane magnets\cite{ezawa2011compact,ezawa2010skyrmion,ezawa2011skyrmion,gobel2019magnetic,gobel2021beyond,kim2019dynamics,yu2018transformation,gao2019creation,ohara2022reversible,zhang2020static,shen2020current,chmiel2018observation,jani2021antiferromagnetic,zhang2021dynamic,zhang2021frustrated,araujo2020typical}.
Rooted in the concept of topological solitons, bimerons are essentially composed of two meron entities coupled together to form a nanometer-sized spin structure. The anisotropic Dzyaloshinsky-Moriya interaction (DMI) plays a key role in the formation of bimeron soliton\cite{gobel2019magnetic,shen2020dynamics}.
However, accessible DMI provided by the non-centrosymmetric system or the ferromagnet/heavy metal interface are both isotropic. In this case, the bimeron solitons are energy-unfavorable and tend to cluster and form assembly or chain-like structures, which are also reported as domain wall bimeron (DWBM)\cite{yu2024spontaneous,li2020bimeron,augustin2021properties,mukai2024polymerization,castro2023skyrmion}.
Recent experimental researches have confirmed the existence of Bloch-type meron chains in the ferromagnetic CoZnMn~\cite{nagase2021observation}, the ferrimagnetic GdFeCo~\cite{li2021field}, and the antiferromagnetic CuMnAs~\cite{amin2023antiferromagnetic}. The N{\'e}el-type version of DWBM has also been found in Van der Waals Ferromagnet FeGeTe~\cite{lv2022controllable,gao2020spontaneous}.

Similar to magnetic skyrmions, the bimerons on the domain wall are topologically protected, offering significant thermal stability suitable for long-term storage. The size of the bimeron is generally constrained by the width of the domain wall, which typically spans tens of nanometers in ferromagnets. Moreover, the domain wall can serve as tracks that host bimerons, showing promise for potential applications in spintronic memory devices. Despite extensive investigations into the stabilization of DWBMs, their dynamics have not yet been reported, and effective methods to manipulate them remain elusive. Notably, the role of the skyrmion Hall effect—an intrinsic characteristic of quasi-particles with nonzero topological charge~\cite{Wang2022NC, Yang2023NC}—has not been explored in the dynamics of DWBMs.

In this study, we employ both analytical and numerical methods to examine the statics and dynamics of Néel-type DWBM driven by spin-orbit torque (SOT). Compared to the isolated bimeron soliton, the thermal stability of DWBM is enhanced by the magnetic domain wall, and it withstands a broad range of DMI strengths. The dynamics excited by the current relate to the polarization direction of the spin current. Our analytical findings suggest that the equivalent force due to magnetic damping is significantly reduced along the domain wall. Consequently, the domain wall can guide the motion of the bimeron and effectively mitigate the skyrmion Hall effect (SkHE) ~\cite{jiang2015blowing,zhang2016magnetic}. More importantly, our results confirm that the spin arrangement of DWBMs allows the skyrmion Hall effect~\cite{jiang2015blowing,zhang2016magnetic} to be harnessed as the primary driving force. Therefore, the dynamics of bimerons on the domain wall can be significantly accelerated, showing an increase in mobility by approximately 40 times compared to several isolated topological solitons, including skyrmions and bimerons, when directly driven by spin-orbit torque.

\begin{figure}[H]
\includegraphics[width=0.8\textwidth]{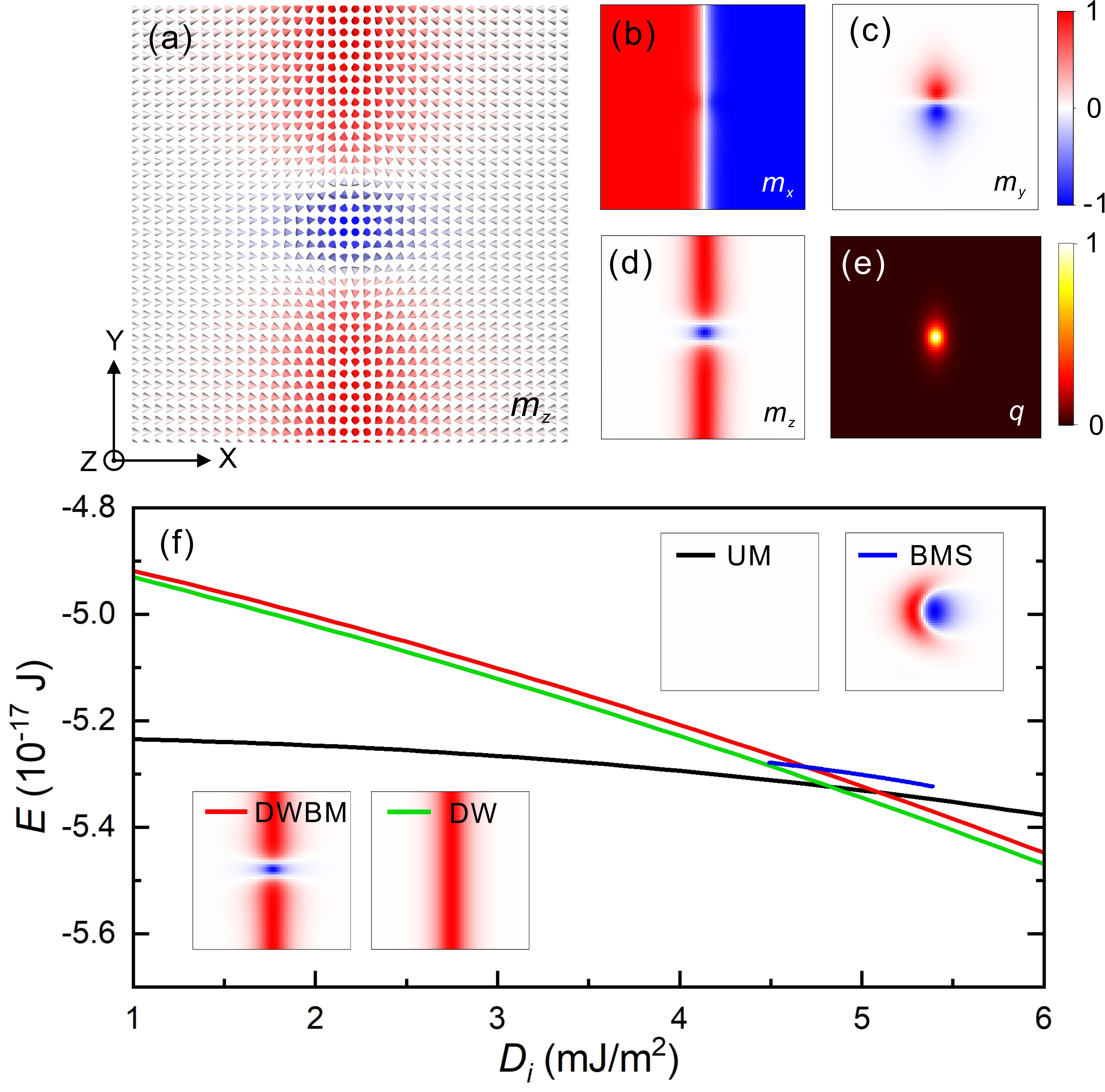}
\caption{Structure and statics of the ferromagnetic DWBM with topological number Q = 1. Distribution of 
(a) spin vectors, (b)-(d) the magnetization components $m_x$, $m_y$, and $m_z$ and the (e) topological charge density, respectively. The color scales represent the quantities noted in bottom right. (f) Total energy of four different magnetic states varies with the Dzyaloshinskii-Moriya interaction constant $D_i$. The energy of the system is recorded after the magnetization is fully relaxed. The inset shows the initial state used for simulation, and the color scale represents the magnetization component $m_z$.}
\label{FIG1}
\end{figure}

Considering a bilayer heterostructure composed of the ferromagnet (FM) layer with in-plane easy axis that hosts DWBM, and the heavy metal (HM) layer serving as the spin Hall channel, while the FM/HM interface provides DMI (Please see Method for the modeling details). Figure~\ref{FIG1}(a) shows the spin structure of the numerically stabilized N{\'e}el-type DWBM, where the bimeron part is characterized by a magnetization transition of $360^\circ$ along the domain wall, and a transition of $180^\circ$ in the direction normal to it. The distribution the magnetization components are shown in Figure~\ref{FIG1}(b)-(d). The magnetic topology of the bimeron is defined by the integer-valued topological charge $Q = \frac{1}{4\pi}{\int{\mathrm{d}x\mathrm{d}y} ~\boldsymbol{m}\cdot(\frac{\partial\boldsymbol{m}}{\partial{x}}\times\frac{\partial\boldsymbol{m}}{\partial{y}})}$, and the topological charge density is shown in Figure~\ref{FIG1}(e). We note that the isolated bimeron soliton can be stabilized in the same system. The isotropic DMI breaks the symmetry of its spin structure and introduces nonreciprocal dynamics excited by spin current and spin waves\cite{liang2023bidirectional,shen2022nonreciprocal,babu2023tunable,jin2022spin}. We involved this particular excitation in our later discussion.

Figure~\ref{FIG1}(f) shows the free energy of varied spin textures as functions of interfacial DMI constant $D_i$. We use the uniformed magnetization (UM), N{\'e}el-type domain wall (DW), bimeron soliton (BMS), and domain wall bimeron (DWBM) as the initial states of the simulation, as shown in the inset. Both the energies of DW (green line) and DWBM (red line) decrease with $D_i$. The critical DMI constant is about 5 mJ/m$^2$, above which UM becomes energy-unfavorable. Despite an energy penalty over the DW, the DWBM can be stabilized within a wide range of $D_i$ from 1 mJ/m$^2$ to 6 mJ/m$^2$. We note that the N{\'e}el-type domain wall plays a role of the stable ground state with relatively high energy, locally elevates the energy barrier of the DWBM, and subsequently improves its stability. A detailed stabilization phase diagram by measuring the sizes of DWBM is shown in Supporting Information Figure~S1. In comparison, the BMS (blue line) survives within a small range of $D_i$ from 4.4 mJ/m$^2$ to 5.3 mJ/m$^2$.


For the investigation of the DWBM dynamics, we combine numerical and analytical approaches to obtain the motion velocity of DWBM driven by currents with the spin polarization direction in X and Y, respectively. The velocities of the bimeron are defined as $v_{i} = \dot{r}_{i}$, and the guiding center ($r_{x}$, $r_{y}$) is numerically calculated by 
\begin{equation}
r_{i}=\frac {\int{\mathrm{d}x\mathrm{d}y}\left[i\boldsymbol{m}\cdot\left(\frac{\partial\boldsymbol{m}}{\partial{x}}\times\frac{\partial\boldsymbol{m}}{\partial{y}}\right)\right]} {\int{\mathrm{d}x\mathrm{d}y}\left[\boldsymbol{m}\cdot\left(\frac{\partial\boldsymbol{m}}{\partial{x}}\times\frac{\partial\boldsymbol{m}}{\partial{y}}\right)\right]},\quad i = x,y.\tag{1}
\label{eq:1}
\end{equation}
%
%
 We assume collective motion of the bimeron with the domain wall, so $v_x$ also denotes the speed of domain wall. On the other hand, the dynamics of the DWBM can be analytically explained by considering the balance between the equivalent forces, which is also known as the Thiele's or the collective coordinate approach\cite{thiele1973steady,tveten2013staggered,tretiakov2008dynamics,clarke2008dynamics}. The steady motion of the bimeron and the domain wall requires   
\begin{equation}
\boldsymbol{G}\times\boldsymbol{v}-\alpha\mathcal{D}\boldsymbol{v} -4\pi \mathcal{C}\hat{p} = \boldsymbol{0}.\tag{2}
\label{eq:2}
\end{equation}
Here the Magnus force is $\boldsymbol{F}_\text{G} = \boldsymbol{G}\times\boldsymbol{v}$ and $\boldsymbol{G} = -4\pi Q \boldsymbol{e}_z$ is the gyrovector. 
The damping force is $\boldsymbol{F}_{\alpha} = -\alpha\mathcal{D}\boldsymbol{v}$ and $\mathcal{D} = 4\pi [C_{ij}]$ represents the dissipative force tensor, in which $i,j$ stands for the coordinate $x,y$ and $D_{ij} = 1/4\pi \int d_{ij}~\mathrm{d}x\mathrm{d}y $ with the components of the dissipative force density tensor $d_{ij} = \partial_{i}\boldsymbol{m}\cdot\partial_{j}\boldsymbol{m}$. 
The driving force arises from the damping-like spin-orbit torque, characterized by the expression $\boldsymbol{F}_\text{SOT} = -4\pi \mathcal{C}\hat{p}$, where $\mathcal{C}$ represents the driving force tensor defined as $\mathcal{C} = \tau{_\text{SH}} [C_{ij}]$. And, $C_{ij} = 1/4\pi \int c_{ij}~\mathrm{d}x\mathrm{d}y$, with the components of the driving force density tensor given by $c_{ij} = (\partial_i\boldsymbol{m}\times\boldsymbol{m})_j$, where the subscript $j$ denotes the $j$th component of the vector $\partial_i\boldsymbol{m}\times\boldsymbol{m}$. The unit current polarization vector $\hat{p}$ is specified as $\boldsymbol{e}_x$ or $\boldsymbol{e}_y$.
Ultimately, Eq.~(\ref{eq:2}) can be reduced to obtain a linear equation for velocity of steady motion in the film plane (Please refer to the Supporting Information for details), 
\begin{equation}
    \begin{bmatrix}
      \alpha D_{xx} & \alpha D_{xy} -Q \\
      \alpha D_{yx} + Q & \alpha D_{yy}
    \end{bmatrix}
    \begin{bmatrix} 
        v_x\\
        v_y
    \end{bmatrix} = -\tau_\mathrm{SH}
    \begin{bmatrix} 
        C_{xx}&C_{xy}\\
        C_{yx}&C_{yy}\\
    \end{bmatrix}
    \begin{bmatrix} 
        p_x\\
        p_y
    \end{bmatrix} 
.\tag{3}
\label{eq:3}
\end{equation}
%
\begin{figure}[t]
\centerline{\includegraphics[width=0.8\textwidth]{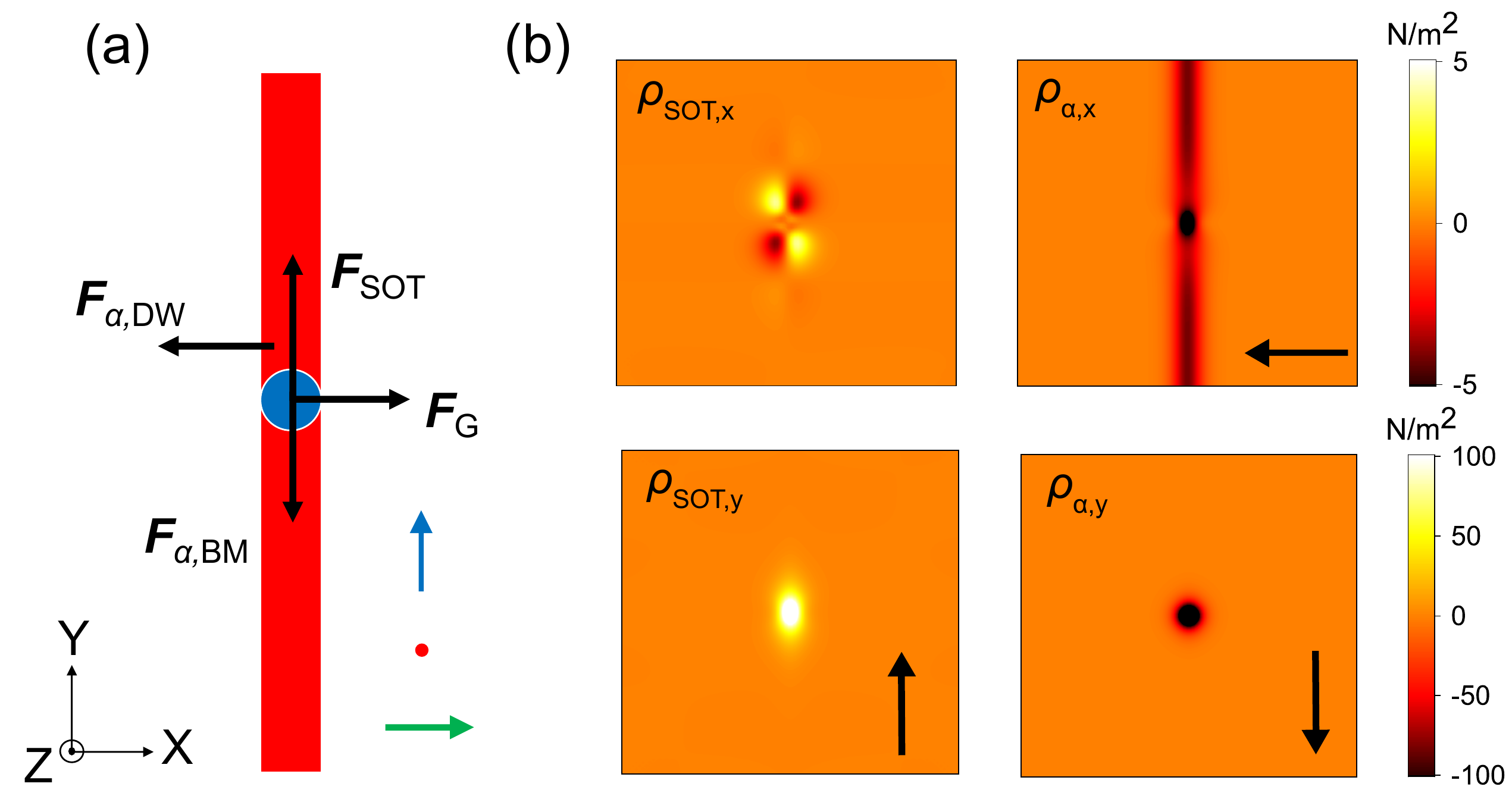}}
\caption{Dynamic mechanism of DWBM driven by spin current with X polarization. (a) Schematic diagram of the equivalent forces involved, including the damping force of the domain wall $\boldsymbol{F}_{\alpha,\text{DW}}$, and the bimeron $\boldsymbol{F}_{\alpha,\text{BM}}$, the Magnus force $\boldsymbol{F}_G$ and the driving force from SOT $\boldsymbol{F}_{\text{SOT}}$. The red part and the blue part denote the domain wall and the bimeron, respectively. The blue and green arrows in the right corner represent the direction of the DWBM motion and the current polarization, while the red dot indicates the domain wall is static. (b) Distribution of the force density components of SOT, $\rho_{\text{SOT},i}$, and magnetic damping, $\rho_{\alpha,i}$, with the applied current density $j_c$= 10$^{10}$A/m$^2$. The black arrows indicate the direction of the resultant force.}
\label{FIG2}
\end{figure}

\begin{figure}[t]
\centerline{\includegraphics[width=1\textwidth]{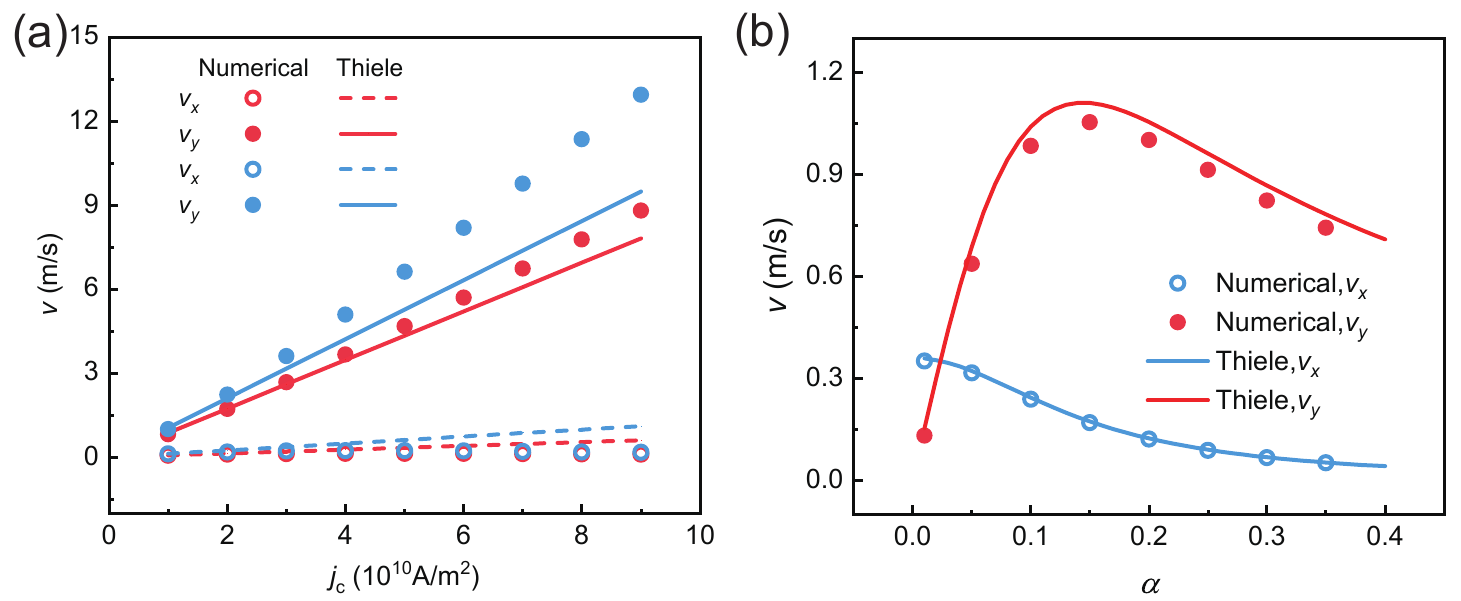}}
\caption{Bimeron velocities introduced by spin current with X polarization. (a) Velocities as functions of current density $j_c$ with damping constant $\alpha = 0.3$ (red) and $0.2$ (blue). (b) Velocities as functions of $\alpha$, with $j_c$= 10$^{10}$A/m$^2$.}
\label{FIG3}
\end{figure}

Our initial investigation focused on studying the dynamics of the DWBM driven by spin current polarized in the X direction. In this scenario, the charge current in the heavy metal layer flows in the direction of the domain wall. Figure~\ref{FIG2}(a) shows the schematic diagram of the forces involved. The SOT contributes to a net equivalent force in the +Y direction, which competes the magnetic damping of the bimeron.
On the other hand, both the bimeron and the domain wall contribute to the damping force in -X, resulting in very low mobility in this direction. 
Considering a system with high damping, the Magnus force of the bimeron could be sufficiently compensated, and thus suppressing the skyrmion Hall effect. Figure~\ref{FIG2}(b) shows the corresponding force density distribution when a current with $j_c=10^{10}$A/m$^2$ is applied. Micromagnetic simulation confirms the above-mentioned mechanism, as shown in Figure~\ref{FIG3}. We observed the linear dependency between the velocity $v_x$, $v_y$ and the charge current density $j_c$, as predicted by Eq.~(\ref{eq:3}). Moreover, the translational speed $v_x$ is very small, indicating that the skyrmion Hall effect is effectively mitigated (Please refer to the Supporting Information, Movie S1). Since the N{\'e}el-type magnetic domain wall is topologically trivial, the Magnus force locally affects the bimeron and tends to bend the magnetic domain wall gradually. The bending of the domain wall is more obvious in the cases with low damping and high current density, leading to the deviation in velocity shown in Figure~\ref{FIG3}(a). Figure~\ref{FIG3}(b) shows the bimeron velocity as a function of the damping constant $\alpha$ when $j_c = 10^{10}$ A/m$^{2}$. The skyrmion Hall effect observed in the low-damping regime ($\alpha < 0.2$) depends on the specific domain wall length used for the analyses. In general, as long as the magnetic domain wall can provide sufficient magnetic damping and elastic restoring, the skyrmion Hall effect of the bimeron can be well controlled. We note that a recent experimental research demonstrates a very similar dynamic mechanism of the domain wall anti-skyrmion in systems with perpendicular magnetic anisotropy~\cite{He2024NM}. 

\begin{figure}[t]
\centerline{\includegraphics[width=0.8\textwidth]{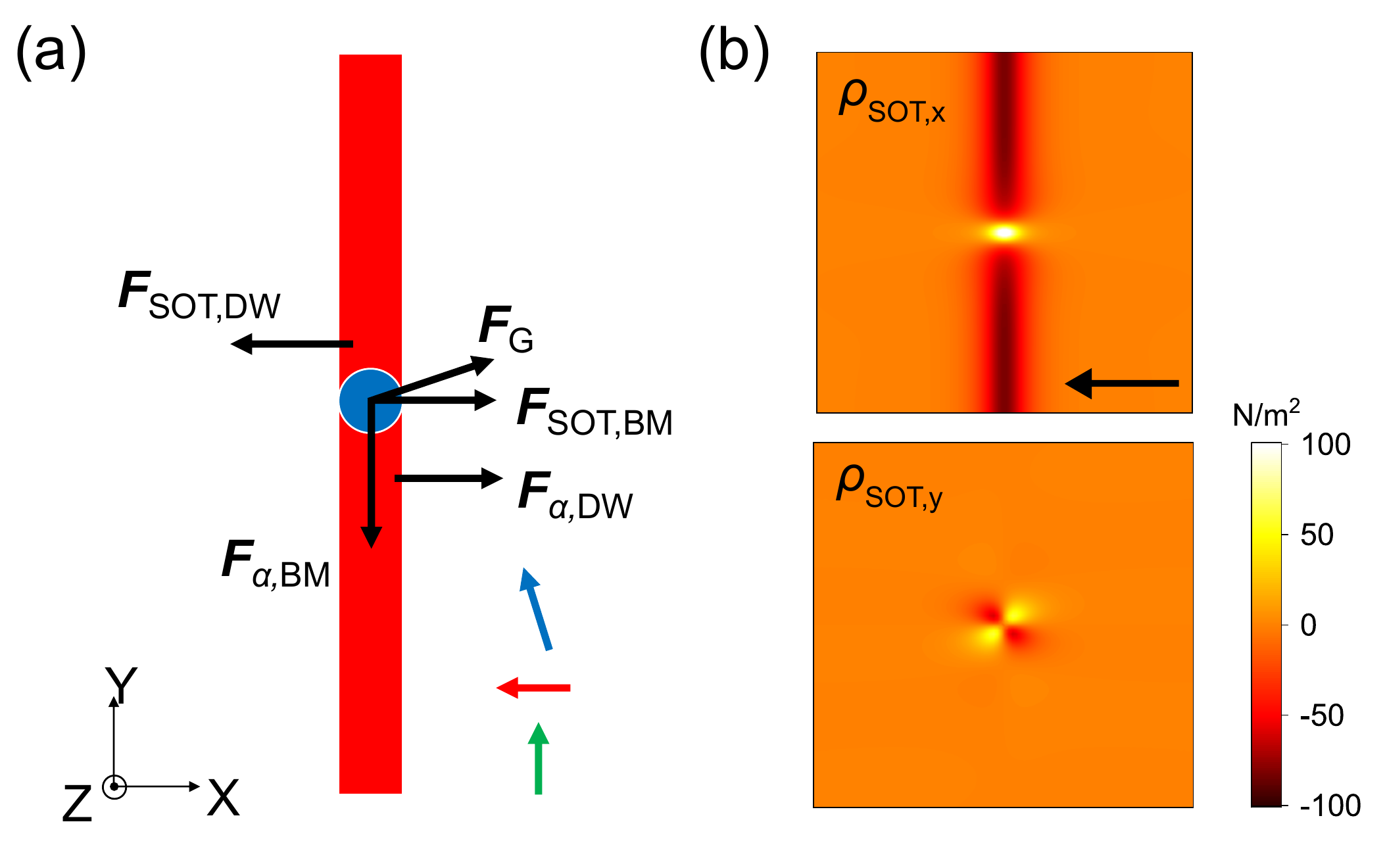}}
\caption{Dynamic mechanism of DWBM driven by the spin current with Y polarization. (a) Schematic diagram of the equivalent forces involved. Note that the SOT forces for the domain wall and DWBM have opposite directions. The blue, red, and green arrows in the right corner demonstrate the direction of the DWBM motion, the domain wall motion, and the current polarization, respectively. (b) Distribution of the force density components of SOT with the applied current density $j_c$= 10$^{10}$A/m$^2$. The damping force densities remain the same with Figure~\ref{FIG2}(b).}
\label{FIG4}
\end{figure}

\begin{figure}[H]
\centerline{\includegraphics[width=\textwidth]{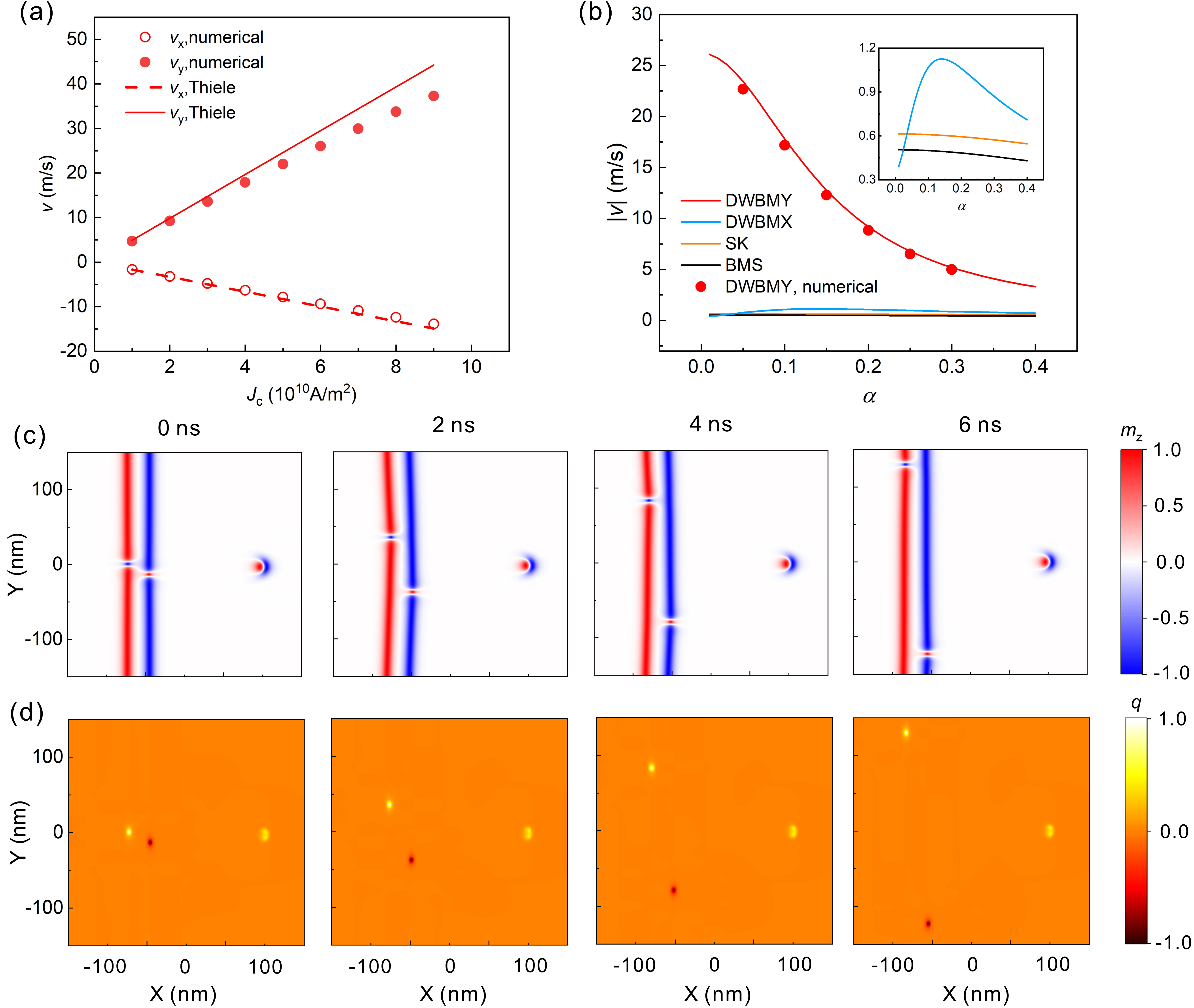}}
\caption{DWBM velocities introduced by spin current with Y polarization. (a) Velocities as functions of current density $j_c$ with damping constant $\alpha = 0.3$. $v_y$ manifests the skyrmion Hall effect of the bimeron. (b) Speed $|v|$ of domain wall bimeron driven by current with X (DWBMX) and Y (DWBMY) polarization, isolated bimeron soliton (BMS), and skyrmion (SK), as functions of $\alpha$, with $j_c$= 10$^{10}$A/m$^2$. For the simulation of skyrmion, we used perpendicular magnetic anisotropy and a DMI constant of 3 mJ/m$^2$, while the other parameters are the same as that of DWBM. The inset zooms in the bottom part of figure to show the speed of spin textures directly driven by SOT. Snapshots of (c) magnetization component $m_z$ and (d) topological charge density $q$ when current with $j_c$= 10$^{10}$A/m$^2$ is applied. The damping constant is set at 0.05. 
}
\label{FIG5}
\end{figure}

When the charge current is polarized in the Y direction, the dynamics of the DWBM are fundamentally different from the above-discussed mechanism. Figure~\ref{FIG4}(a) shows the schematic diagram of the forces involved in this scenario. The SOT provides the driving forces in the -X and +X direction for the domain wall and the bimeron, respectively. The corresponding force density distribution is shown in Figure~\ref{FIG4}(b). 
Considering the dimension of the domain wall is much larger than the bimeron, the dynamics of the domain wall will dominate, and the SOT drives the spin textures into motion in the -X direction. In this case, the bimeron is not directly driven by the SOT, but instead follows the collective motion of the domain wall, leading to a strong Magnus force component in the orthogonal direction (in our case the +Y direction). The Magnus force component competes with the magnetic damping of the bimeron and results in a very high velocity. Briefly, the dynamics of the bimeron is boosted by the skyrmion Hall effect. We note this mechanism is entirely different from that of skyrmion or bimeron soliton, for which the Magnus force always joins the magnetic damping force to compete with the spin-orbit torque, and negatively impacts their mobility. Micromagnetic simulations demonstrate the linear relationship between the current density $j_c$ and the velocity of the bimeron, and we observed a significant improvement in the mobility, which is also confirmed by the analytical results obtained by Eq.~(\ref{eq:3}), as shown in Figure~\ref{FIG5}(a). We further compared the speed $|v|$ of the DWBM, isolated bimeron soliton (BMS), and skyrmion (SK) driven by SOT, as shown in Figure~\ref{FIG5}(b). The mobility of DWBM is more than an order larger than that of SK or BMS and non-linearly decreases with $\alpha$. This is because the magnetic damping force competes with the driving force of domain wall from SOT, and the driving force of bimeron from skyrmion Hall effect. By decreasing $\alpha$ to 0.05, both numerical simulation and analytical approach confirm a 37 times increase in the mobility of DWBM compared with SK, and a 49 times increase compared with BMS.
Figure~\ref{FIG5}(c) visualizes this difference by showing the displacements of DWBMs and the BMS within the same system introduced by spin current (Please refer to the Supporting Information, Movie S2), and Figure~\ref{FIG5}(d) shows the corresponding distribution of topological charge density $q$. Over a period of 6 ns, BMS remains nearly static, while DWBMs with opposite topological numbers travel fast in opposite directions, demonstrating the skyrmion Hall effect-based sliding motion. In addition, it is worth noting that when directly excited by SOT, the driving force scales with the size of topological spin textures\cite{shen2019current,rohart2013skyrmion}, while the Magnus force remains constant due to the topological charge conservation. For compact spin textures with size of nanometers, the dynamics will be dominated by the skyrmion Hall effect and the efficiency of spin-orbit torque will significantly decrease. On the other hand, if it is possible to utilize the skyrmion Hall effect as a driving force, as we presented in the scenario of DWBM, higher mobility is achievable. This will in turn benefit the power consumption of DWBM-based spintronic devices.

In conclusion, we analytically and numerically study the statics of domain wall bimerons, and their dynamics induced by damping-like spin-orbit torque. The numerical simulations show that the domain wall bimeron can be stabilized within a wide range of DMI, and the stability is effectively improved compared with the bimeron soliton.
The motion of domain wall bimeron induced by spin-orbit torque is also discussed, and speed is analytically derived, which agrees well with the numerical simulations. When the current is applied in the direction of the domain wall, the skyrmion Hall effect of the domain wall bimeron can be effectively suppressed by the damping effect of the magnetic domain wall. More importantly, when the current is applied in the direction perpendicular to the domain wall, the skyrmion Hall effect serves as the primary driving force and significantly boosts the motion velocity of the domain wall bimeron. Both the numerical simulation and analytical equation demonstrate approximately 40 times increase in the mobility of the domain wall bimeron compared with skyrmion and bimeron soliton in ferromagnetic systems with a damping constant of 0.05. Our results demonstrate unique dynamics of domain wall bimeron related to skyrmion Hall effect, and provide effective ways for building bimeron-based spintronic devices.

\section{Method}
\noindent\textbf{Numerical Simulations.} MuMax~\cite{vansteenkiste2014design} is used to perform the micromagnetic simulations. Considering a bilayer heterostructure composed of the ferromagnetic thin film that host DWBM, and the heavy metal layer serving as the spin Hall channel. The free energy of the FM film can be described by
\begin{equation}
\begin{aligned}
E=&\int{\mathrm{d}V}  \left\{A_{ex}(\nabla\boldsymbol{m})^{\text{2}}-K_u(\boldsymbol{m}\cdot\boldsymbol{n})^{2}-\frac{1}{2}\mu_{0} M_{\text{S}} \boldsymbol{H_d}\cdot\boldsymbol{m} 
+D_i[m_{z}(\nabla\cdot\boldsymbol{m})-(\boldsymbol{m}\cdot\nabla)m_{z}] \right\},
\end{aligned}
\tag{4}
\label{eq:4}
\end{equation}
where the first, second, third, and fourth terms correspond to the energy density of Heisenberg exchange, magnetic anisotropy, dipolar field, and interfacial Dzyaloshinskii-Moriya interaction, respectively.
$A_{ex}$, $K_u$ and $D_i$ are the exchange constant, magnetic anisotropy constant and DMI constant.
$\boldsymbol{n}=\boldsymbol{e}_y$ stands for the direction of the anisotropy easy axis, and $\boldsymbol{H_d}$ is the dipole field. 

We employ the Landau-Lifshitz-Gilbert (LLG) equation~\cite{gilbert2004phenomenological} with the damping-like spin-orbit torque to simulate the dynamics of FM systems, which is described as
\begin{equation}
\frac{\partial \boldsymbol{m}}{\partial t} = -\gamma \boldsymbol{m} \times \boldsymbol{H}_{\text{eff}} + \alpha \boldsymbol{m} \times \frac{\partial \boldsymbol{m}}{\partial t} - \tau_\mathrm{SH} \boldsymbol{m} \times (\boldsymbol{m} \times \hat{p}),
\tag{5}
\label{eq:5}
\end{equation}
where $\boldsymbol{m}$ is the normalized magnetization. 
The damping-like spin-orbit torque $- \tau_\mathrm{SH} \boldsymbol{m} \times (\boldsymbol{m} \times \hat{p})$ can be produced by the spin Hall effect of the adjacent HM layer. 
Here $\hat{p}$ is the polarization vector and the magnitude of the damping-like torque is defined as $\tau_{\mathrm{SH}} = \frac{\gamma\hbar j_c\theta_{\mathrm{SH}}}{2\mu_0etM_\mathrm{S}}$ with the current density $j_c$, the reduced Planck constant $\hbar$, the spin Hall angle $\theta_{\text{SH}}$, the vacuum permeability constant $\mu_{0}$, the elementary charge $e$, and the layer thickness $t$. 
$\gamma$ and $\alpha$ denote the gyromagnetic ratio and the damping constant respectively.
$\boldsymbol{H_{\text{eff}}}$ stands for the effective field obtained from the variation of the FM energy in Eq. (\ref{eq:4}). We assumed the Rashba effect is weak compared with the spin Hall effect in the FM/HM bilayer system, and excluded the influence from field-like spin-orbit torque.
For the micromagnetic simulations, we employed the parameters of CoFeB/Pt~\cite{litzius2017skyrmion}, while the DMI constant and damping constant were varied within a reasonable range: Exchange constant $A_{ex}$ = 15 pJ/m, anisotropy coefficient $K_{u}$ = 0.8 MJ/m$^3$, saturated magnetization $M_{\mathrm{S}}$ = 580 kA/m, DMI strength $D_i$ = 1 mJ/m$^2$ to 6 mJ/m$^2$, damping constant $\alpha$ = 0.05 to 0.4, gyromagnetic ratio $\gamma$ = 2.211 $\times 10^5$ m/(A$\cdot$s) and spin Hall angle $\theta_\text{SH}$ = 0.1. The mesh size of 1024 $\times$ 1024 $\times$ 1 nm$^3$ is used.

\section{Associated Content}
\textbf{Data Availability Statement}

\noindent The data that support the findings of this study are available
from the corresponding authors upon reasonable request. The
micromagnetic simulations are performed using the freely
available MuMax3 platform, which is publicly accessible at
https://mumax.github.io.

\noindent\textbf{Supporting Information}


\begin{itemize}
\item[] The details regarding the derivation of Thiele's equation for the steady motion velocity of domain wall bimeron, and the phase diagram for the size of domain wall bimeron stabilized with varied magnetic anisotropy and DMI strength. (PDF)

\item[] Movie S1: The motion of the domain wall bimerons with topological charge $Q$ = +1 and -1, and a bimeron soliton with topological charge $Q$ = +1 introduced by spin current polarized in +X direction. The total simulation time is 30ns. The skyrmion Hall effect of the domain wall bimeron is suppressed, while that of the bimeron soliton is demonstrated by the translational drift. (AVI)

\item[] Movie S2: The motion of the domain wall bimerons with topological charge $Q$ = +1 and -1, and a bimeron soliton with topological charge $Q$ = +1 introduced by spin current polarized in +Y direction. The total simulation time is 7ns. The sliding motion of domain wall bimerons is driven by skyrmion Hall effect. (AVI)
\end{itemize}

\section{Author Information}

\noindent\textbf{Corresponding Authors}

Xiaoguang Li; Email: lixiaoguang@sztu.edu.cn

Yan Zhou; Email: zhouyan@cuhk.edu.cn


\begin{acknowledgement}


The authors thank funding from the National Key R\&D Program of China (Grant no. 2022YFA1603200, 2022YFA1603202), the National Natural Science Foundation of China (Grant No. 12104322, No. 12375237), the Shenzhen Science and Technology Program (Grant No. ZDSYS20200811143600001), the National Natural Science Foundation of China (Grant No. 52001215), Guangdong Basic and Applied Basic Research Foundation (Grant No. 2021A1515012055), the industrial research and development project of SZTU (Grant No. KY2022QJKCZ005), the Shenzhen Peacock Group Plan (KQTD20180413181702403), the Shenzhen Fundamental Research Fund (Grant No. JCYJ20210324120213037), the Guangdong Basic and Applied Basic Research Foundation (2021B1515120047), the National Natural Science Foundation of China (12374123). X.Z. acknowledges support by CREST, the Japan Science and Technology Agency (Grant No. JPMJCR20T1).

\end{acknowledgement}

\bibliography{mybib}

\providecommand{\latin}[1]{#1}
\makeatletter
\providecommand{\doi}
  {\begingroup\let\do\@makeother\dospecials
  \catcode`\{=1 \catcode`\}=2 \doi@aux}
\providecommand{\doi@aux}[1]{\endgroup\texttt{#1}}
\makeatother
\providecommand*\mcitethebibliography{\thebibliography}
\csname @ifundefined\endcsname{endmcitethebibliography}  {\let\endmcitethebibliography\endthebibliography}{}
\begin{mcitethebibliography}{46}
\providecommand*\natexlab[1]{#1}
\providecommand*\mciteSetBstSublistMode[1]{}
\providecommand*\mciteSetBstMaxWidthForm[2]{}
\providecommand*\mciteBstWouldAddEndPuncttrue
  {\def\EndOfBibitem{\unskip.}}
\providecommand*\mciteBstWouldAddEndPunctfalse
  {\let\EndOfBibitem\relax}
\providecommand*\mciteSetBstMidEndSepPunct[3]{}
\providecommand*\mciteSetBstSublistLabelBeginEnd[3]{}
\providecommand*\EndOfBibitem{}
\mciteSetBstSublistMode{f}
\mciteSetBstMaxWidthForm{subitem}{(\alph{mcitesubitemcount})}
\mciteSetBstSublistLabelBeginEnd
  {\mcitemaxwidthsubitemform\space}
  {\relax}
  {\relax}

\bibitem[Ezawa(2011)]{ezawa2011compact}
Ezawa,~M. Compact merons and skyrmions in thin chiral magnetic films. \emph{Phys. Rev. B} \textbf{2011}, \emph{83}, 100408\relax
\mciteBstWouldAddEndPuncttrue
\mciteSetBstMidEndSepPunct{\mcitedefaultmidpunct}
{\mcitedefaultendpunct}{\mcitedefaultseppunct}\relax
\EndOfBibitem
\bibitem[Ezawa and Tsitsishvili(2010)Ezawa, and Tsitsishvili]{ezawa2010skyrmion}
Ezawa,~Z.; Tsitsishvili,~G. Skyrmion and bimeron excitations in bilayer quantum Hall systems. \emph{Physica E} \textbf{2010}, \emph{42}, 1069--1072\relax
\mciteBstWouldAddEndPuncttrue
\mciteSetBstMidEndSepPunct{\mcitedefaultmidpunct}
{\mcitedefaultendpunct}{\mcitedefaultseppunct}\relax
\EndOfBibitem
\bibitem[Ezawa and Tsitsishvili(2011)Ezawa, and Tsitsishvili]{ezawa2011skyrmion}
Ezawa,~Z.; Tsitsishvili,~G. Skyrmion and bimeron excitations in imbalanced bilayer quantum Hall systems. AIP Conference Proceedings. 2011; pp 605--606\relax
\mciteBstWouldAddEndPuncttrue
\mciteSetBstMidEndSepPunct{\mcitedefaultmidpunct}
{\mcitedefaultendpunct}{\mcitedefaultseppunct}\relax
\EndOfBibitem
\bibitem[G{\"o}bel \latin{et~al.}(2019)G{\"o}bel, Mook, Henk, Mertig, and Tretiakov]{gobel2019magnetic}
G{\"o}bel,~B.; Mook,~A.; Henk,~J.; Mertig,~I.; Tretiakov,~O.~A. Magnetic bimerons as skyrmion analogues in in-plane magnets. \emph{Phys. Rev. B} \textbf{2019}, \emph{99}, 060407\relax
\mciteBstWouldAddEndPuncttrue
\mciteSetBstMidEndSepPunct{\mcitedefaultmidpunct}
{\mcitedefaultendpunct}{\mcitedefaultseppunct}\relax
\EndOfBibitem
\bibitem[G{\"o}bel \latin{et~al.}(2021)G{\"o}bel, Mertig, and Tretiakov]{gobel2021beyond}
G{\"o}bel,~B.; Mertig,~I.; Tretiakov,~O.~A. Beyond skyrmions: Review and perspectives of alternative magnetic quasiparticles. \emph{Phys. Rep.} \textbf{2021}, \emph{895}, 1--28\relax
\mciteBstWouldAddEndPuncttrue
\mciteSetBstMidEndSepPunct{\mcitedefaultmidpunct}
{\mcitedefaultendpunct}{\mcitedefaultseppunct}\relax
\EndOfBibitem
\bibitem[Kim(2019)]{kim2019dynamics}
Kim,~S.~K. Dynamics of bimeron skyrmions in easy-plane magnets induced by a spin supercurrent. \emph{Phys. Rev. B} \textbf{2019}, \emph{99}, 224406\relax
\mciteBstWouldAddEndPuncttrue
\mciteSetBstMidEndSepPunct{\mcitedefaultmidpunct}
{\mcitedefaultendpunct}{\mcitedefaultseppunct}\relax
\EndOfBibitem
\bibitem[Yu \latin{et~al.}(2018)Yu, Koshibae, Tokunaga, Shibata, Taguchi, Nagaosa, and Tokura]{yu2018transformation}
Yu,~X.; Koshibae,~W.; Tokunaga,~Y.; Shibata,~K.; Taguchi,~Y.; Nagaosa,~N.; Tokura,~Y. Transformation between meron and skyrmion topological spin textures in a chiral magnet. \emph{Nature} \textbf{2018}, \emph{564}, 95--98\relax
\mciteBstWouldAddEndPuncttrue
\mciteSetBstMidEndSepPunct{\mcitedefaultmidpunct}
{\mcitedefaultendpunct}{\mcitedefaultseppunct}\relax
\EndOfBibitem
\bibitem[Gao \latin{et~al.}(2019)Gao, Je, Im, Choi, Yang, Li, Wang, Lee, Han, Lee, \latin{et~al.} others]{gao2019creation}
Gao,~N.; Je,~S.-G.; Im,~M.-Y.; Choi,~J.~W.; Yang,~M.; Li,~Q.-c.; Wang,~T.; Lee,~S.; Han,~H.-S.; Lee,~K.-S.; others Creation and annihilation of topological meron pairs in in-plane magnetized films. \emph{Nat. Commun.} \textbf{2019}, \emph{10}, 5603\relax
\mciteBstWouldAddEndPuncttrue
\mciteSetBstMidEndSepPunct{\mcitedefaultmidpunct}
{\mcitedefaultendpunct}{\mcitedefaultseppunct}\relax
\EndOfBibitem
\bibitem[Ohara \latin{et~al.}(2022)Ohara, Zhang, Chen, Kato, Xia, Ezawa, Tretiakov, Hou, Zhou, Zhao, \latin{et~al.} others]{ohara2022reversible}
Ohara,~K.; Zhang,~X.; Chen,~Y.; Kato,~S.; Xia,~J.; Ezawa,~M.; Tretiakov,~O.~A.; Hou,~Z.; Zhou,~Y.; Zhao,~G.; others Reversible Transformation between Isolated Skyrmions and Bimerons. \emph{Nano Lett.} \textbf{2022}, \emph{22}, 8559--8566\relax
\mciteBstWouldAddEndPuncttrue
\mciteSetBstMidEndSepPunct{\mcitedefaultmidpunct}
{\mcitedefaultendpunct}{\mcitedefaultseppunct}\relax
\EndOfBibitem
\bibitem[Zhang \latin{et~al.}(2020)Zhang, Xia, Shen, Ezawa, Tretiakov, Zhao, Liu, and Zhou]{zhang2020static}
Zhang,~X.; Xia,~J.; Shen,~L.; Ezawa,~M.; Tretiakov,~O.~A.; Zhao,~G.; Liu,~X.; Zhou,~Y. Static and dynamic properties of bimerons in a frustrated ferromagnetic monolayer. \emph{Phys. Rev. B} \textbf{2020}, \emph{101}, 144435\relax
\mciteBstWouldAddEndPuncttrue
\mciteSetBstMidEndSepPunct{\mcitedefaultmidpunct}
{\mcitedefaultendpunct}{\mcitedefaultseppunct}\relax
\EndOfBibitem
\bibitem[Shen \latin{et~al.}(2020)Shen, Xia, Zhang, Ezawa, Tretiakov, Liu, Zhao, and Zhou]{shen2020current}
Shen,~L.; Xia,~J.; Zhang,~X.; Ezawa,~M.; Tretiakov,~O.~A.; Liu,~X.; Zhao,~G.; Zhou,~Y. Current-Induced Dynamics and Chaos of Antiferromagnetic Bimerons. \emph{Phys. Rev. Lett.} \textbf{2020}, \emph{124}, 037202\relax
\mciteBstWouldAddEndPuncttrue
\mciteSetBstMidEndSepPunct{\mcitedefaultmidpunct}
{\mcitedefaultendpunct}{\mcitedefaultseppunct}\relax
\EndOfBibitem
\bibitem[Chmiel \latin{et~al.}(2018)Chmiel, Waterfield~Price, Johnson, Lamirand, Schad, van~der Laan, Harris, Irwin, Rzchowski, Eom, \latin{et~al.} others]{chmiel2018observation}
Chmiel,~F.~P.; Waterfield~Price,~N.; Johnson,~R.~D.; Lamirand,~A.~D.; Schad,~J.; van~der Laan,~G.; Harris,~D.~T.; Irwin,~J.; Rzchowski,~M.~S.; Eom,~C.-B.; others Observation of magnetic vortex pairs at room temperature in a planar $\alpha$-Fe2O3/Co heterostructure. \emph{Nat. Mater.} \textbf{2018}, \emph{17}, 581--585\relax
\mciteBstWouldAddEndPuncttrue
\mciteSetBstMidEndSepPunct{\mcitedefaultmidpunct}
{\mcitedefaultendpunct}{\mcitedefaultseppunct}\relax
\EndOfBibitem
\bibitem[Jani \latin{et~al.}(2021)Jani, Lin, Chen, Harrison, Maccherozzi, Schad, Prakash, Eom, Ariando, Venkatesan, \latin{et~al.} others]{jani2021antiferromagnetic}
Jani,~H.; Lin,~J.-C.; Chen,~J.; Harrison,~J.; Maccherozzi,~F.; Schad,~J.; Prakash,~S.; Eom,~C.-B.; Ariando,~A.; Venkatesan,~T.; others Antiferromagnetic half-skyrmions and bimerons at room temperature. \emph{Nature} \textbf{2021}, \emph{590}, 74--79\relax
\mciteBstWouldAddEndPuncttrue
\mciteSetBstMidEndSepPunct{\mcitedefaultmidpunct}
{\mcitedefaultendpunct}{\mcitedefaultseppunct}\relax
\EndOfBibitem
\bibitem[Zhang \latin{et~al.}(2021)Zhang, Xia, Tretiakov, Diep, Zhao, Yang, Zhou, Ezawa, and Liu]{zhang2021dynamic}
Zhang,~X.; Xia,~J.; Tretiakov,~O.~A.; Diep,~H.~T.; Zhao,~G.; Yang,~J.; Zhou,~Y.; Ezawa,~M.; Liu,~X. Dynamic transformation between a skyrmion string and a bimeron string in a layered frustrated system. \emph{Phys. Rev. B} \textbf{2021}, \emph{104}, L220406\relax
\mciteBstWouldAddEndPuncttrue
\mciteSetBstMidEndSepPunct{\mcitedefaultmidpunct}
{\mcitedefaultendpunct}{\mcitedefaultseppunct}\relax
\EndOfBibitem
\bibitem[Zhang \latin{et~al.}(2021)Zhang, Xia, Ezawa, Tretiakov, Diep, Zhao, Liu, and Zhou]{zhang2021frustrated}
Zhang,~X.; Xia,~J.; Ezawa,~M.; Tretiakov,~O.~A.; Diep,~H.~T.; Zhao,~G.; Liu,~X.; Zhou,~Y. A frustrated bimeronium: Static structure and dynamics. \emph{Appl. Phys. Lett.} \textbf{2021}, \emph{118}\relax
\mciteBstWouldAddEndPuncttrue
\mciteSetBstMidEndSepPunct{\mcitedefaultmidpunct}
{\mcitedefaultendpunct}{\mcitedefaultseppunct}\relax
\EndOfBibitem
\bibitem[Ara{\'u}jo \latin{et~al.}(2020)Ara{\'u}jo, Lopes, Carvalho-Santos, Pereira, Silva, Silva, and Altbir]{araujo2020typical}
Ara{\'u}jo,~A.; Lopes,~R.; Carvalho-Santos,~V.; Pereira,~A.; Silva,~R.; Silva,~R.; Altbir,~D. Typical skyrmions versus bimerons: A long-distance competition in ferromagnetic racetracks. \emph{Phys. Rev. B} \textbf{2020}, \emph{102}, 104409\relax
\mciteBstWouldAddEndPuncttrue
\mciteSetBstMidEndSepPunct{\mcitedefaultmidpunct}
{\mcitedefaultendpunct}{\mcitedefaultseppunct}\relax
\EndOfBibitem
\bibitem[Shen \latin{et~al.}(2020)Shen, Li, Xia, Qiu, Zhang, Tretiakov, Ezawa, and Zhou]{shen2020dynamics}
Shen,~L.; Li,~X.; Xia,~J.; Qiu,~L.; Zhang,~X.; Tretiakov,~O.~A.; Ezawa,~M.; Zhou,~Y. Dynamics of ferromagnetic bimerons driven by spin currents and magnetic fields. \emph{Phys. Rev. B} \textbf{2020}, \emph{102}, 104427\relax
\mciteBstWouldAddEndPuncttrue
\mciteSetBstMidEndSepPunct{\mcitedefaultmidpunct}
{\mcitedefaultendpunct}{\mcitedefaultseppunct}\relax
\EndOfBibitem
\bibitem[Yu \latin{et~al.}(2024)Yu, Kanazawa, Zhang, Takahashi, Iakoubovskii, Nakajima, Tanigaki, Mochizuki, and Tokura]{yu2024spontaneous}
Yu,~X.; Kanazawa,~N.; Zhang,~X.; Takahashi,~Y.; Iakoubovskii,~K.~V.; Nakajima,~K.; Tanigaki,~T.; Mochizuki,~M.; Tokura,~Y. Spontaneous Vortex-Antivortex Pairs and Their Topological Transitions in a Chiral-Lattice Magnet. \emph{Adv. Mater.} \textbf{2024}, \emph{36}, 2306441\relax
\mciteBstWouldAddEndPuncttrue
\mciteSetBstMidEndSepPunct{\mcitedefaultmidpunct}
{\mcitedefaultendpunct}{\mcitedefaultseppunct}\relax
\EndOfBibitem
\bibitem[Li \latin{et~al.}(2020)Li, Shen, Bai, Wang, Zhang, Xia, Ezawa, Tretiakov, Xu, Mruczkiewicz, \latin{et~al.} others]{li2020bimeron}
Li,~X.; Shen,~L.; Bai,~Y.; Wang,~J.; Zhang,~X.; Xia,~J.; Ezawa,~M.; Tretiakov,~O.~A.; Xu,~X.; Mruczkiewicz,~M.; others Bimeron clusters in chiral antiferromagnets. \emph{npj Comput. Mater.} \textbf{2020}, \emph{6}, 169\relax
\mciteBstWouldAddEndPuncttrue
\mciteSetBstMidEndSepPunct{\mcitedefaultmidpunct}
{\mcitedefaultendpunct}{\mcitedefaultseppunct}\relax
\EndOfBibitem
\bibitem[Augustin \latin{et~al.}(2021)Augustin, Jenkins, Evans, Novoselov, and Santos]{augustin2021properties}
Augustin,~M.; Jenkins,~S.; Evans,~R.~F.; Novoselov,~K.~S.; Santos,~E.~J. Properties and dynamics of meron topological spin textures in the two-dimensional magnet CrCl3. \emph{Nat. Commun.} \textbf{2021}, \emph{12}, 185\relax
\mciteBstWouldAddEndPuncttrue
\mciteSetBstMidEndSepPunct{\mcitedefaultmidpunct}
{\mcitedefaultendpunct}{\mcitedefaultseppunct}\relax
\EndOfBibitem
\bibitem[Mukai and Leonov(2024)Mukai, and Leonov]{mukai2024polymerization}
Mukai,~N.; Leonov,~A.~O. “Polymerization” of Bimerons in Quasi-Two-Dimensional Chiral Magnets with Easy-Plane Anisotropy. \emph{Nanomaterials} \textbf{2024}, \emph{14}, 504\relax
\mciteBstWouldAddEndPuncttrue
\mciteSetBstMidEndSepPunct{\mcitedefaultmidpunct}
{\mcitedefaultendpunct}{\mcitedefaultseppunct}\relax
\EndOfBibitem
\bibitem[Castro \latin{et~al.}(2023)Castro, Altbir, Galvez-Poblete, Corona, Oyarz{\'u}n, Pereira, Allende, and Carvalho-Santos]{castro2023skyrmion}
Castro,~M.; Altbir,~D.; Galvez-Poblete,~D.; Corona,~R.; Oyarz{\'u}n,~S.; Pereira,~A.; Allende,~S.; Carvalho-Santos,~V. Skyrmion-bimeron dynamic conversion in magnetic nanotracks. \emph{Phys. Rev. B} \textbf{2023}, \emph{108}, 094436\relax
\mciteBstWouldAddEndPuncttrue
\mciteSetBstMidEndSepPunct{\mcitedefaultmidpunct}
{\mcitedefaultendpunct}{\mcitedefaultseppunct}\relax
\EndOfBibitem
\bibitem[Nagase \latin{et~al.}(2021)Nagase, So, Yasui, Ishida, Yoshida, Tanaka, Saitoh, Ikarashi, Kawaguchi, Kuwahara, \latin{et~al.} others]{nagase2021observation}
Nagase,~T.; So,~Y.-G.; Yasui,~H.; Ishida,~T.; Yoshida,~H.~K.; Tanaka,~Y.; Saitoh,~K.; Ikarashi,~N.; Kawaguchi,~Y.; Kuwahara,~M.; others Observation of domain wall bimerons in chiral magnets. \emph{Nat. Commun.} \textbf{2021}, \emph{12}, 3490\relax
\mciteBstWouldAddEndPuncttrue
\mciteSetBstMidEndSepPunct{\mcitedefaultmidpunct}
{\mcitedefaultendpunct}{\mcitedefaultseppunct}\relax
\EndOfBibitem
\bibitem[Li \latin{et~al.}(2021)Li, Su, Lin, Liu, Gao, Wang, Wei, Zhao, Zhang, Cai, \latin{et~al.} others]{li2021field}
Li,~Z.; Su,~J.; Lin,~S.-Z.; Liu,~D.; Gao,~Y.; Wang,~S.; Wei,~H.; Zhao,~T.; Zhang,~Y.; Cai,~J.; others Field-free topological behavior in the magnetic domain wall of ferrimagnetic GdFeCo. \emph{Nat. Commun.} \textbf{2021}, \emph{12}, 5604\relax
\mciteBstWouldAddEndPuncttrue
\mciteSetBstMidEndSepPunct{\mcitedefaultmidpunct}
{\mcitedefaultendpunct}{\mcitedefaultseppunct}\relax
\EndOfBibitem
\bibitem[Amin \latin{et~al.}(2023)Amin, Poole, Reimers, Barton, Dal~Din, Maccherozzi, Dhesi, Nov{\'a}k, Krizek, Chauhan, \latin{et~al.} others]{amin2023antiferromagnetic}
Amin,~O.; Poole,~S.; Reimers,~S.; Barton,~L.; Dal~Din,~A.; Maccherozzi,~F.; Dhesi,~S.; Nov{\'a}k,~V.; Krizek,~F.; Chauhan,~J.; others Antiferromagnetic half-skyrmions electrically generated and controlled at room temperature. \emph{Nat. Nanotechnol.} \textbf{2023}, \emph{18}, 849--853\relax
\mciteBstWouldAddEndPuncttrue
\mciteSetBstMidEndSepPunct{\mcitedefaultmidpunct}
{\mcitedefaultendpunct}{\mcitedefaultseppunct}\relax
\EndOfBibitem
\bibitem[Lv \latin{et~al.}(2022)Lv, Pei, Yang, Qin, Liu, Zhang, and Che]{lv2022controllable}
Lv,~X.; Pei,~K.; Yang,~C.; Qin,~G.; Liu,~M.; Zhang,~J.; Che,~R. Controllable topological magnetic transformations in the thickness-tunable van der Waals ferromagnet Fe5GeTe2. \emph{ACS Nano} \textbf{2022}, \emph{16}, 19319--19327\relax
\mciteBstWouldAddEndPuncttrue
\mciteSetBstMidEndSepPunct{\mcitedefaultmidpunct}
{\mcitedefaultendpunct}{\mcitedefaultseppunct}\relax
\EndOfBibitem
\bibitem[Gao \latin{et~al.}(2020)Gao, Yin, Wang, Li, Cai, Zhao, Lei, Wang, Zhang, and Shen]{gao2020spontaneous}
Gao,~Y.; Yin,~Q.; Wang,~Q.; Li,~Z.; Cai,~J.; Zhao,~T.; Lei,~H.; Wang,~S.; Zhang,~Y.; Shen,~B. Spontaneous (anti) meron chains in the domain walls of van der Waals ferromagnetic Fe5- xGeTe2. \emph{Adv. Mater.} \textbf{2020}, \emph{32}, 2005228\relax
\mciteBstWouldAddEndPuncttrue
\mciteSetBstMidEndSepPunct{\mcitedefaultmidpunct}
{\mcitedefaultendpunct}{\mcitedefaultseppunct}\relax
\EndOfBibitem
\bibitem[Wang \latin{et~al.}(2022)Wang, Song, Wei, Nan, Zhang, Ge, Tian, Zang, and Du]{Wang2022NC}
Wang,~W.; Song,~D.; Wei,~W.~W.; Nan,~P.~N.; Zhang,~S.; Ge,~B.; Tian,~M.; Zang,~J.; Du,~H. Electrical manipulation of skyrmions in a chiral magnet. \emph{Nat. Commun.} \textbf{2022}, \emph{13}, 1593\relax
\mciteBstWouldAddEndPuncttrue
\mciteSetBstMidEndSepPunct{\mcitedefaultmidpunct}
{\mcitedefaultendpunct}{\mcitedefaultseppunct}\relax
\EndOfBibitem
\bibitem[Yang \latin{et~al.}(2023)Yang, Zhao, Wu, Chu, Xu, Li, Åkerman, and Zhou]{Yang2023NC}
Yang,~S.; Zhao,~Y.; Wu,~K.; Chu,~Z.; Xu,~X.; Li,~X.; Åkerman,~J.; Zhou,~Y. Reversible conversion between skyrmions and skyrmioniums. \emph{Nat. Commun.} \textbf{2023}, \emph{14}, 3406\relax
\mciteBstWouldAddEndPuncttrue
\mciteSetBstMidEndSepPunct{\mcitedefaultmidpunct}
{\mcitedefaultendpunct}{\mcitedefaultseppunct}\relax
\EndOfBibitem
\bibitem[Jiang \latin{et~al.}(2015)Jiang, Upadhyaya, Zhang, Yu, Jungfleisch, Fradin, Pearson, Tserkovnyak, Wang, Heinonen, \latin{et~al.} others]{jiang2015blowing}
Jiang,~W.; Upadhyaya,~P.; Zhang,~W.; Yu,~G.; Jungfleisch,~M.~B.; Fradin,~F.~Y.; Pearson,~J.~E.; Tserkovnyak,~Y.; Wang,~K.~L.; Heinonen,~O.; others Blowing magnetic skyrmion bubbles. \emph{Science} \textbf{2015}, \emph{349}, 283--286\relax
\mciteBstWouldAddEndPuncttrue
\mciteSetBstMidEndSepPunct{\mcitedefaultmidpunct}
{\mcitedefaultendpunct}{\mcitedefaultseppunct}\relax
\EndOfBibitem
\bibitem[Zhang \latin{et~al.}(2016)Zhang, Zhou, and Ezawa]{zhang2016magnetic}
Zhang,~X.; Zhou,~Y.; Ezawa,~M. Magnetic bilayer-skyrmions without skyrmion Hall effect. \emph{Nat. Commun.} \textbf{2016}, \emph{7}, 10293\relax
\mciteBstWouldAddEndPuncttrue
\mciteSetBstMidEndSepPunct{\mcitedefaultmidpunct}
{\mcitedefaultendpunct}{\mcitedefaultseppunct}\relax
\EndOfBibitem
\bibitem[Liang \latin{et~al.}(2023)Liang, Lan, Zhao, Zelent, Krawczyk, and Zhou]{liang2023bidirectional}
Liang,~X.; Lan,~J.; Zhao,~G.; Zelent,~M.; Krawczyk,~M.; Zhou,~Y. Bidirectional magnon-driven bimeron motion in ferromagnets. \emph{Phys. Rev. B} \textbf{2023}, \emph{108}, 184407\relax
\mciteBstWouldAddEndPuncttrue
\mciteSetBstMidEndSepPunct{\mcitedefaultmidpunct}
{\mcitedefaultendpunct}{\mcitedefaultseppunct}\relax
\EndOfBibitem
\bibitem[Shen \latin{et~al.}(2022)Shen, Xia, Chen, Li, Zhang, Tretiakov, Shao, Zhao, Liu, Ezawa, \latin{et~al.} others]{shen2022nonreciprocal}
Shen,~L.; Xia,~J.; Chen,~Z.; Li,~X.; Zhang,~X.; Tretiakov,~O.~A.; Shao,~Q.; Zhao,~G.; Liu,~X.; Ezawa,~M.; others Nonreciprocal dynamics of ferrimagnetic bimerons. \emph{Phys. Rev. B} \textbf{2022}, \emph{105}, 014422\relax
\mciteBstWouldAddEndPuncttrue
\mciteSetBstMidEndSepPunct{\mcitedefaultmidpunct}
{\mcitedefaultendpunct}{\mcitedefaultseppunct}\relax
\EndOfBibitem
\bibitem[Babu \latin{et~al.}(2023)Babu, Perumal, Sankaran~Kunnath, and Sinha]{babu2023tunable}
Babu,~P.; Perumal,~H.~P.; Sankaran~Kunnath,~S.; Sinha,~J. Tunable Creation and Annihilation of Magnetic Bimerons in Square-Shaped Submicron Dot. \emph{ACS Appl. Elec. Mater.} \textbf{2023}, \emph{6}, 221--229\relax
\mciteBstWouldAddEndPuncttrue
\mciteSetBstMidEndSepPunct{\mcitedefaultmidpunct}
{\mcitedefaultendpunct}{\mcitedefaultseppunct}\relax
\EndOfBibitem
\bibitem[Jin \latin{et~al.}(2022)Jin, Li, Zhang, Wang, Wang, Lian, Gong, and Shi]{jin2022spin}
Jin,~C.; Li,~S.; Zhang,~H.; Wang,~R.; Wang,~J.; Lian,~R.; Gong,~P.; Shi,~X. Spin-wave modes of magnetic bimerons in nanodots. \emph{New J. Phys.} \textbf{2022}, \emph{24}, 073013\relax
\mciteBstWouldAddEndPuncttrue
\mciteSetBstMidEndSepPunct{\mcitedefaultmidpunct}
{\mcitedefaultendpunct}{\mcitedefaultseppunct}\relax
\EndOfBibitem
\bibitem[Thiele(1973)]{thiele1973steady}
Thiele,~A. Steady-State Motion of Magnetic Domains. \emph{Phys. Rev. Lett.} \textbf{1973}, \emph{30}, 230\relax
\mciteBstWouldAddEndPuncttrue
\mciteSetBstMidEndSepPunct{\mcitedefaultmidpunct}
{\mcitedefaultendpunct}{\mcitedefaultseppunct}\relax
\EndOfBibitem
\bibitem[Tveten \latin{et~al.}(2013)Tveten, Qaiumzadeh, Tretiakov, and Brataas]{tveten2013staggered}
Tveten,~E.~G.; Qaiumzadeh,~A.; Tretiakov,~O.; Brataas,~A. Staggered Dynamics in Antiferromagnets by Collective Coordinates. \emph{Phys. Rev. Lett.} \textbf{2013}, \emph{110}, 127208\relax
\mciteBstWouldAddEndPuncttrue
\mciteSetBstMidEndSepPunct{\mcitedefaultmidpunct}
{\mcitedefaultendpunct}{\mcitedefaultseppunct}\relax
\EndOfBibitem
\bibitem[Tretiakov \latin{et~al.}(2008)Tretiakov, Clarke, Chern, Bazaliy, and Tchernyshyov]{tretiakov2008dynamics}
Tretiakov,~O.; Clarke,~D.; Chern,~G.-W.; Bazaliy,~Y.~B.; Tchernyshyov,~O. Dynamics of Domain Walls in Magnetic Nanostrips. \emph{Phys. Rev. Lett.} \textbf{2008}, \emph{100}, 127204\relax
\mciteBstWouldAddEndPuncttrue
\mciteSetBstMidEndSepPunct{\mcitedefaultmidpunct}
{\mcitedefaultendpunct}{\mcitedefaultseppunct}\relax
\EndOfBibitem
\bibitem[Clarke \latin{et~al.}(2008)Clarke, Tretiakov, Chern, Bazaliy, and Tchernyshyov]{clarke2008dynamics}
Clarke,~D.; Tretiakov,~O.; Chern,~G.-W.; Bazaliy,~Y.~B.; Tchernyshyov,~O. Dynamics of a vortex domain wall in a magnetic nanostrip: Application of the collective-coordinate approach. \emph{Phys. Rev. B} \textbf{2008}, \emph{78}, 134412\relax
\mciteBstWouldAddEndPuncttrue
\mciteSetBstMidEndSepPunct{\mcitedefaultmidpunct}
{\mcitedefaultendpunct}{\mcitedefaultseppunct}\relax
\EndOfBibitem
\bibitem[He \latin{et~al.}(2024)He, Li, Chen, Wang, Shen, Wang, Song, Zhao, Lin, Zhang, and Shen]{He2024NM}
He,~Z.; Li,~Z.; Chen,~Z.; Wang,~Z.; Shen,~J.; Wang,~S.; Song,~C.; Zhao,~J.,~Tongyun~Cai; Lin,~S.-Z.; Zhang,~Y.; Shen,~B. Experimental observation of current-driven antiskyrmion sliding in stripe domains. \emph{Nat. Mater.} \textbf{2024}, 1--7\relax
\mciteBstWouldAddEndPuncttrue
\mciteSetBstMidEndSepPunct{\mcitedefaultmidpunct}
{\mcitedefaultendpunct}{\mcitedefaultseppunct}\relax
\EndOfBibitem
\bibitem[Shen \latin{et~al.}(2019)Shen, Li, Zhao, Xia, Zhao, and Zhou]{shen2019current}
Shen,~L.; Li,~X.; Zhao,~Y.; Xia,~J.; Zhao,~G.; Zhou,~Y. Current-Induced Dynamics of the Antiferromagnetic Skyrmion and Skyrmionium. \emph{Phys. Rev. Appl.} \textbf{2019}, \emph{12}, 064033\relax
\mciteBstWouldAddEndPuncttrue
\mciteSetBstMidEndSepPunct{\mcitedefaultmidpunct}
{\mcitedefaultendpunct}{\mcitedefaultseppunct}\relax
\EndOfBibitem
\bibitem[Rohart and Thiaville(2013)Rohart, and Thiaville]{rohart2013skyrmion}
Rohart,~S.; Thiaville,~A. Skyrmion confinement in ultrathin film nanostructures in the presence of Dzyaloshinskii-Moriya interaction. \emph{Phys. Rev. B} \textbf{2013}, \emph{88}, 184422\relax
\mciteBstWouldAddEndPuncttrue
\mciteSetBstMidEndSepPunct{\mcitedefaultmidpunct}
{\mcitedefaultendpunct}{\mcitedefaultseppunct}\relax
\EndOfBibitem
\bibitem[Vansteenkiste \latin{et~al.}(2014)Vansteenkiste, Leliaert, Dvornik, Helsen, Garcia-Sanchez, and Van~Waeyenberge]{vansteenkiste2014design}
Vansteenkiste,~A.; Leliaert,~J.; Dvornik,~M.; Helsen,~M.; Garcia-Sanchez,~F.; Van~Waeyenberge,~B. The design and verification of MuMax3. \emph{AIP adv.} \textbf{2014}, \emph{4}\relax
\mciteBstWouldAddEndPuncttrue
\mciteSetBstMidEndSepPunct{\mcitedefaultmidpunct}
{\mcitedefaultendpunct}{\mcitedefaultseppunct}\relax
\EndOfBibitem
\bibitem[Gilbert(2004)]{gilbert2004phenomenological}
Gilbert,~T.~L. A phenomenological theory of damping in ferromagnetic materials. \emph{IEEE Trans. Magn.} \textbf{2004}, \emph{40}, 3443--3449\relax
\mciteBstWouldAddEndPuncttrue
\mciteSetBstMidEndSepPunct{\mcitedefaultmidpunct}
{\mcitedefaultendpunct}{\mcitedefaultseppunct}\relax
\EndOfBibitem
\bibitem[Litzius \latin{et~al.}(2017)Litzius, Lemesh, Kr{\"u}ger, Bassirian, Caretta, Richter, B{\"u}ttner, Sato, Tretiakov, F{\"o}rster, \latin{et~al.} others]{litzius2017skyrmion}
Litzius,~K.; Lemesh,~I.; Kr{\"u}ger,~B.; Bassirian,~P.; Caretta,~L.; Richter,~K.; B{\"u}ttner,~F.; Sato,~K.; Tretiakov,~O.~A.; F{\"o}rster,~J.; others Skyrmion Hall effect revealed by direct time-resolved X-ray microscopy. \emph{Nat. Phys.} \textbf{2017}, \emph{13}, 170--175\relax
\mciteBstWouldAddEndPuncttrue
\mciteSetBstMidEndSepPunct{\mcitedefaultmidpunct}
{\mcitedefaultendpunct}{\mcitedefaultseppunct}\relax
\EndOfBibitem
\end{mcitethebibliography}

\end{document}